\pdfoutput=1
\documentclass[prl,letterpaper,twocolumn,10pt,aps,notitlepage,nofootinbib,nobibnotes,superscriptaddress]{revtex4-1}

\usepackage{amsfonts}
\usepackage{amsmath}
\usepackage{bm,graphicx}
\usepackage{hyperref}

\hypersetup{colorlinks, linkcolor = [rgb]{0,0.0,0.75}, citecolor = [rgb]{0,0.0,0.75}, urlcolor = [rgb]{0,0.0,0.75}}

\usepackage[utf8]{inputenc}

\begin{document}

\title{High-precision calculation of the strange \\ nucleon electromagnetic form factors}

\author{Jeremy Green}
\email{green@kph.uni-mainz.de}
  \affiliation{Institut f\"ur Kernphysik, Johannes Gutenberg-Universit\"at Mainz, D-55099 Mainz, Germany}

\author{Stefan Meinel}
\email{smeinel@email.arizona.edu}
  \affiliation{Department of Physics, University of Arizona, Tucson, AZ 85721, USA}
  \affiliation{RIKEN BNL Research Center, Brookhaven National Laboratory, Upton, NY 11973, USA}

\author{Michael Engelhardt}
  \affiliation{Department of Physics, New Mexico State University, Las Cruces, NM 88003-8001, USA}

\author{Stefan Krieg}
  \affiliation{Bergische Universit\"at Wuppertal, D-42119 Wuppertal, Germany}
  \affiliation{IAS, J\"ulich Supercomputing Centre, Forschungszentrum J\"ulich, D-52425 J\"ulich, Germany}

\author{\mbox{Jesse Laeuchli}}
  \affiliation{Department of Computer Science, College of William and Mary, Williamsburg, VA 23187, USA}
  
\author{John Negele}
  \affiliation{Center for Theoretical Physics, Massachusetts Institute of Technology, 
               Cambridge, Massachusetts 02139, USA}

\author{Kostas Orginos}
  \affiliation{Physics Department, College of William and Mary, Williamsburg, VA 23187, USA}
  \affiliation{Thomas Jefferson National Accelerator Facility, Newport News, Virginia 23606, USA}

\author{Andrew Pochinsky}
  \affiliation{Center for Theoretical Physics, Massachusetts Institute of Technology, 
               Cambridge, Massachusetts 02139, USA}

\author{Sergey Syritsyn} 
  \affiliation{RIKEN BNL Research Center, Brookhaven National Laboratory, Upton, NY 11973, USA}

\date{May 22, 2015}

\begin{abstract}
  We report a direct lattice QCD calculation of the strange nucleon
  electromagnetic form factors $G_E^s$ and $G_M^s$ in the kinematic range
  $0 \leq Q^2 \lesssim 1.2\: {\rm GeV}^2$. For the first time, both $G_E^s$ and
  $G_M^s$ are shown to be nonzero with high significance. This work uses
  closer-to-physical lattice parameters than previous calculations, and achieves
  an unprecedented statistical precision by implementing a recently proposed
  variance reduction technique called \emph{hierarchical probing}.
  We perform model-independent fits of the form factor shapes using the
  $z$-expansion and determine the strange electric and magnetic radii
  and magnetic moment. We compare our results to parity-violating
  electron-proton scattering data and to other theoretical studies.
\end{abstract}

\maketitle

The nucleon electromagnetic form factors describe how electric charge
and current are distributed inside protons and neutrons, and are therefore
among the most important observables characterizing these building blocks
of ordinary matter. Because nucleons contain only \emph{up} and \emph{down}
valence quarks, these two quark flavors dominate the electromagnetic
form factors. Isolating the small contributions from the other quark
flavors is a significant challenge for both experiment and theory, but
is of fundamental importance for our understanding of the structure of
protons and neutrons, and of the nonperturbative dynamics of QCD. After
the up and down quarks, strange quarks are expected to give the
next-largest contribution to the electromagnetic form factors.
The cross section of elastic electron-proton scattering used to
extract the form factors is dominated by photon exchange, which probes
the sum of all quark-flavor contributions weighted according to their
electric charges. However, by analyzing the small parity-violating
effects arising from interference with $Z$-boson exchange, the
strange-quark contribution to the electromagnetic form factors can be isolated
\cite{Kaplan:1988ku, Mckeown:1989ir}. The available experimental results,
which focus on momentum transfers $Q^2$ in the vicinity of $0.2\:{\rm GeV}^2$,
are consistent with zero but constrain the relative contribution of the
strange quarks to be within a few percent \cite{Spayde:2003nr, Beise:2004py,
Aniol:2004hp, Maas:2004ta, Maas:2004dh, Aniol:2005zf, Aniol:2005zg,
Armstrong:2005hs, Acha:2006my, Androic:2009aa, Baunack:2009gy, Ahmed:2011vp}.

\begin{figure}
\includegraphics[width=\linewidth]{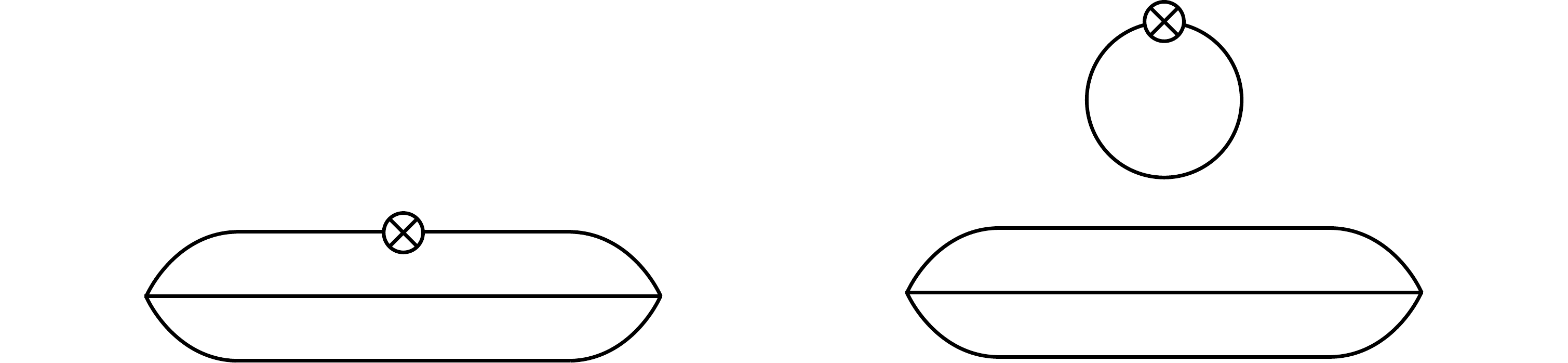}
\caption{\label{fig:CIvsDI}Nonperturbative quark propagator contractions
obtained from Eq.~(\protect\ref{eq:threept}) by performing the path integral
over the quark fields: connected (left) and disconnected (right).}
\end{figure}

Ab-initio calculations of the nucleon electromagnetic form factors $G_E^q$
and $G_M^q$ of an individual quark flavor $q$ (see, e.g.,
Ref.~\cite{Green:2014xba} for the definitions) are possible using
lattice QCD. The form factors can be extracted from
Euclidean three-point functions of the form
\begin{equation}
 \sum_{\mathbf{z},\:\mathbf{y}}e^{-i\mathbf{p}^\prime \cdot (\mathbf{z}-\mathbf{y})}
 e^{-i\mathbf{p} \cdot (\mathbf{y}-\mathbf{x})} \left\langle N_\beta(z) V_q^\mu(y)
 \overline{N}_\alpha(x) \right\rangle, \label{eq:threept}
\end{equation}
where $N$ is an interpolating field with the quantum numbers of the nucleon,
$V_q^\mu = \bar{q}\gamma^\mu q$ is the vector current for quark flavor $q$,
and $\mathbf{p}$, $\mathbf{p^\prime}$ are the spatial momenta of the initial
and final states. In the three-point function (\ref{eq:threept}), performing
the path integral over the quark fields leaves a path integral over the gauge
fields, which contains the product of the fermion determinants and the
nonperturbative quark propagator contractions illustrated in Fig.~\ref{fig:CIvsDI}.
The connected contraction arises only for $q=u,d$ and is numerically large,
while the disconnected contraction is present for all quark flavors
(and is the origin of the strange-quark contribution to the electromagnetic
form factors). The disconnected quark loop in Fig.~\ref{fig:CIvsDI}
has the form
\begin{equation}
 T_q^\mu = -\sum_{\mathbf{y}}e^{i(\mathbf{p}^\prime-\mathbf{p}) \cdot \mathbf{y}}
 \: \mathrm{Tr} \left[ \gamma^\mu D_q^{-1} (y,\:y) \right], \label{eq:loop}
\end{equation}
where $D_q$ is the lattice Dirac operator, and the trace is over color
and spin indices. The numerical computation of the propagator
$D_q^{-1} (y,\:y)$ for all spatial lattice points $\mathbf{y}$ using
standard methods is prohibitively expensive for lattices of realistic size,
and therefore most lattice calculations of nucleon form factors have been
restricted to the connected contractions (and therefore $q=u,d$) only.
In Ref.~\cite{Shanahan:2014tja}, the strange-quark contribution was estimated
by combining experimental data for the complete electromagnetic factors
with lattice QCD results for the connected $u$- and $d$-quark contractions.
This method relies on a delicate cancellation between large quantities,
and is therefore limited in its statistical precision and rather
susceptible to systematic errors in the lattice calculation.

A direct lattice QCD calculation of the strange-quark contribution can be
performed by evaluating the disconnected loop in Eq.~(\ref{eq:loop})
stochastically. An unbiased estimate is given by
\begin{equation}
 T_q^\mu \approx -\frac1N\sum_{n=1}^N\sum_{\mathbf{y}}e^{i(\mathbf{p}^\prime-\mathbf{p})
 \cdot \mathbf{y}} \: \xi_n^\dag(y) \gamma^\mu \psi_n(y), \label{eq:Tnoise}
\end{equation}
where $\xi_n$ are suitable noise vectors (for example with random components
$\in \mathbb{Z}_2$) with support on the time slice $y_0$, and $\psi_n$
are the corresponding solutions of the lattice Dirac equation, $D_q \psi_n = \xi_n$.
Previous unquenched lattice QCD calculations of the strange nucleon
electromagnetic form factors using variants of this approach can be
found in Refs.~\cite{Doi:2009sq} and \cite{Babich:2010at}.
The calculation of Ref.~\cite{Doi:2009sq} was performed on a $16^3\times 32$
lattice with dimensions $(1.9\:{\rm fm})^3\times (3.9\:{\rm fm})$,
and at rather heavy $u$, $d$ quark masses corresponding to pion masses
in the range $600\text{-}840$ MeV. For these parameters, the strange
magnetic form factor $G_M^s$ was found to be negative with a significance
of $2\text{-}3 \sigma$ in the region $Q^2 \lesssim 1\:\:{\rm GeV}^2$,
while the strange electric form factor $G_E^s$ was found to be
consistent with zero \cite{Doi:2009sq}. The authors of Ref.~\cite{Babich:2010at}
used an anisotropic $24^3\times 64$ lattice with dimensions
$(2.6\:{\rm fm})^3\times (2.3\:{\rm fm})$ and a pion mass of 416 MeV.
The corresponding results for $G_M^s$ and $G_E^s$ have large uncertainties
and are consistent with zero.

\begin{figure}
\includegraphics[width=\linewidth]{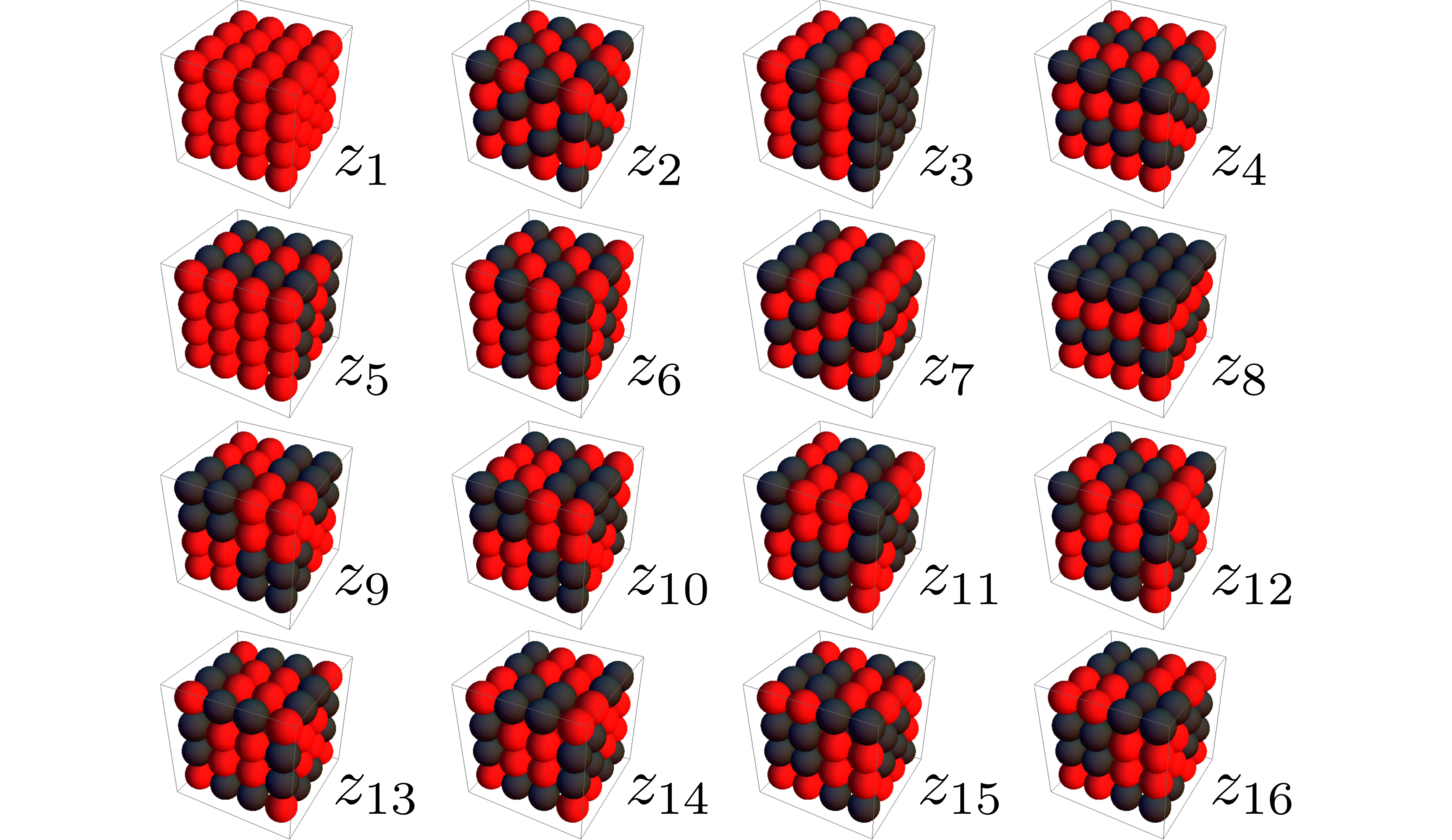}
\caption{\label{fig:hvec}The first 16 Hadamard vectors for hierarchical probing on a $4^3$ lattice; red = $+1$, black = $-1$.}
\end{figure}

In this Letter, we report a new direct lattice QCD calculation of
$G_M^s$ and $G_E^s$ with closer-to-physical parameters and with very high
statistical precision. For the first time, both $G_M^s$ and $G_E^s$ are
shown to be nonzero with high significance and in a wide range of $Q^2$.
We also obtain similarly precise results for the analogous
disconnected light (equal for up and down) quark contributions,
which need to be included in precision lattice calculations of the
total electromagnetic form factors of the proton, and are therefore of
relevance for the proton charge radius puzzle \cite{Pohl:2010zza, Antognini:1900ns}.
Our calculation is performed on a large $32^3 \times 96$ lattice of dimensions
$(3.6\:{\rm fm})^3\times (10.9\:{\rm fm})$ and includes 2+1 flavors of
dynamical sea quarks, implemented using a clover-improved Wilson action.
The up and down quark mass corresponds to a pion mass of $317\:\:{\rm MeV}$,
and the strange-quark mass is consistent with the physical value
(determined using the ``$\eta_s$'' mass \cite{Dowdall:2011wh}) within 5 percent.
The unprecedented statistical precision is achieved as follows:
(i) we use 1028 gauge-field configurations and compute the three-point
function in Eq.~(\ref{eq:threept}) for 96 different source locations,
$x$, on each configuration, and (ii), we use a novel variance reduction
method, \emph{hierarchical probing} \cite{Stathopoulos:2013aci}, to
evaluate the disconnected quark loops $T_q^\mu$. Similarly to \emph{dilution}
\cite{Wilcox:1999ab, Foley:2005ac}, this method is based on the observation
that the fluctuations in Eq.~(\ref{eq:Tnoise}) due to the random noise
vectors originate from the off-diagonal elements of $D^{-1}_q(x,y)$,
which decay with the Euclidean distance $|x-y|$. By partitioning the
noise vectors into multiple vectors with support only on subsets of the
lattice sites, the variance can be reduced (at the cost of additional
solutions of the lattice Dirac equation). While dilution is based on a
fixed partitioning scheme, hierarchical probing allows one to continuously
increase the level of partitioning, eliminating the variance in order
of importance while reusing the results from prior levels \cite{Stathopoulos:2013aci}.
This is achieved using a special sequence of Hadamard vectors,
$z_n$, which have values $\pm1$ on the lattice sites. Examples of $z_n$
are shown in Fig.~\ref{fig:hvec}. In Eq.~(\ref{eq:Tnoise}),
we make the replacement
\begin{equation}
 \xi_n \to z_n \odot \xi,
\end{equation}
where $\xi$ is a single noise vector and $\odot$ denotes the element-wise
Hadamard product \cite{Stathopoulos:2013aci}. As mentioned earlier, we use
noise vectors with support only on selected time slices, and we therefore
perform three-dimensional Hierarchical probing. We use $N=128$ Hadamard
vectors, which eliminates the variance from neighboring lattice sites up
to distance 4. For the electromagnetic form factors, we observe a variance
reduction by approximately a factor of 10 compared to the traditional noise
method with the same $N$ (at equal computational cost). Our calculation
employs complex $\mathbb{Z}_2 \times \mathbb{Z}_2$ noise and also uses
color and spin dilution for the noise vectors $\xi$.

\begin{figure*}
  \includegraphics[width=0.495\linewidth]{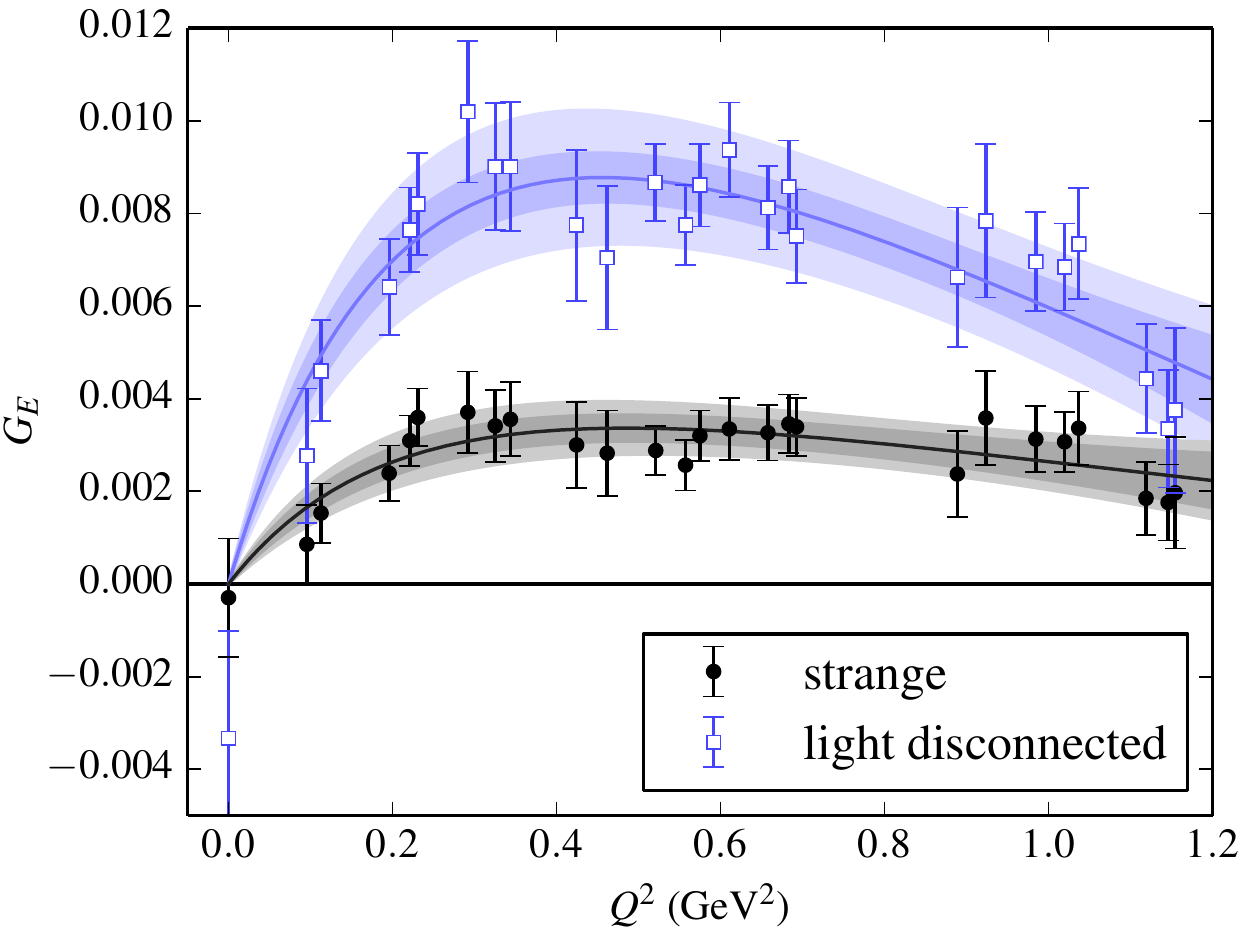}
  \includegraphics[width=0.495\linewidth]{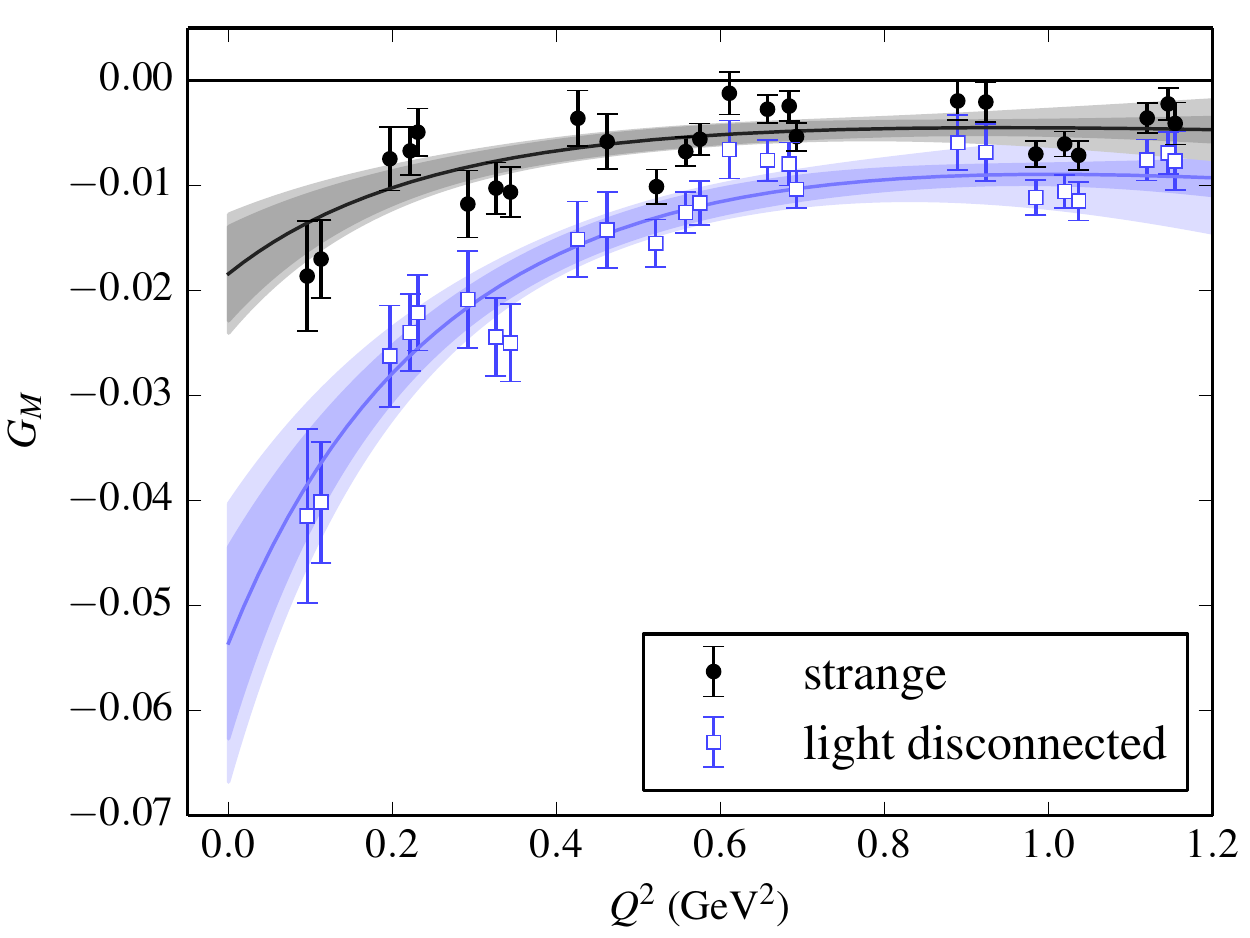}
  \caption{Strange-quark and disconnected light-quark electric and
    magnetic form factors, with statistical error bars. The curves
    result from the $z$-expansion fits; the inner bands show the
    statistical uncertainty and the outer error bands show the
    combined statistical and systematic uncertainties (added in
    quadrature). The charge factors are not included.}
  \label{fig:GE_GM}
\end{figure*}

We use the lattice conserved vector current [generalizing
Eqs.~(\ref{eq:loop}, \ref{eq:Tnoise}) for currents that are not
site-local] including an improvement
term~\cite{Martinelli:1990ny,Boinepalli:2006xd} to reduce $O(a)$
lattice artifacts, where $a$ is the lattice spacing, with the
improvement coefficient set to its tree-level value. We independently
vary the momenta $\mathbf{p}$ and $\mathbf{p}^\prime$ to obtain 59 different values of
$Q^2$. For clarity in plots we bin the nearest points that have the same
$(\mathbf{p}^\prime-\mathbf{p})^2$ and $|\mathbf{p}^{\prime 2}-\mathbf{p}^2|$,
and thus differ in $Q^2$ by less than $0.006\:\:\text{GeV}^2$.
The three-point function (\ref{eq:threept}) receives
contributions from the desired nucleon ground state and from excited states;
the excited-state contributions decay exponentially faster with the source-sink
separation, $|z_0-x_0|$, than the ground-state contribution. We use
$|z_0-x_0|\approx 1.14~{\rm fm}$ for our main results, and estimate the remaining
excited-state contamination using a second separation of $|z_0-x_0|\approx 1.37~{\rm fm}$.

Our calculated form factors are shown in
Fig.~\ref{fig:GE_GM}. $G_E^s(Q^2)$ is consistent with zero at $Q^2=0$
and positive for all other values of $Q^2$. It rises with $Q^2$ until
it reaches a maximum value around 0.003, somewhere between 0.2 and
1.0~GeV$^2$, above which the data hint at a decrease. In the same
range, $G_M^s(Q^2)$ is negative, with a decrease in magnitude with
$Q^2$ up to around 0.5~GeV$^2$. Above that, the data are consistent
with a constant. Note that data with the same spatial momentum
transfer $\mathbf{q}=\mathbf{p}-\mathbf{p}^\prime$ tend to have strongly correlated errors, meaning
that, e.g., the points below $Q^2=0.4$~GeV$^2$ form three clusters of
correlated data that should not be interpreted as ``bumpy'' behavior
in the form factor. The disconnected light-quark form factors have
similar dependence on $Q^2$ but their magnitude is two to three times
that of the strange-quark form factors.

We fit the $Q^2$-dependence of each form factor using the
$z$-expansion~\cite{Hill:2010yb,Epstein:2014zua},
\begin{equation}
  \label{eq:z_exp}
  G(Q^2)=\sum_k^{k_{\text{max}}}a_kz^k,\quad
z=\frac{\sqrt{t_{\text{cut}}+Q^2}-\sqrt{t_{\text{cut}}}}{\sqrt{t_{\text{cut}}+Q^2}+\sqrt{t_{\text{cut}}}},
\end{equation}
which conformally maps the complex domain of analyticity in $Q^2$ to
$|z|<1$. Although for physical quark masses the isoscalar threshold is
$t_{\text{cut}}=(3m_\pi)^2$, at our pion mass we expect that the
$\omega$ resonance is a stable particle below threshold; therefore, we
use the isovector threshold $t_{\text{cut}}=(2m_\pi)^2$ in our
fits. The intercept and slope of the form factor at $Q^2=0$ can be
obtained from the first two coefficients, $a_0$ (which we fix to zero
for $G_E$) and $a_1$. We impose Gaussian priors on the remaining
coefficients, centered at zero with width equal to
$5\max\{|a_0|,|a_1|\}$. We truncate the series with
$k_{\text{max}}=5$, but we have verified that using
$k_{\text{max}}=10$ produces identical fit results in our probed range
of $Q^2$. The resulting fit curves are shown in Fig.~\ref{fig:GE_GM}.

\begin{figure}
  \includegraphics[width=\linewidth]{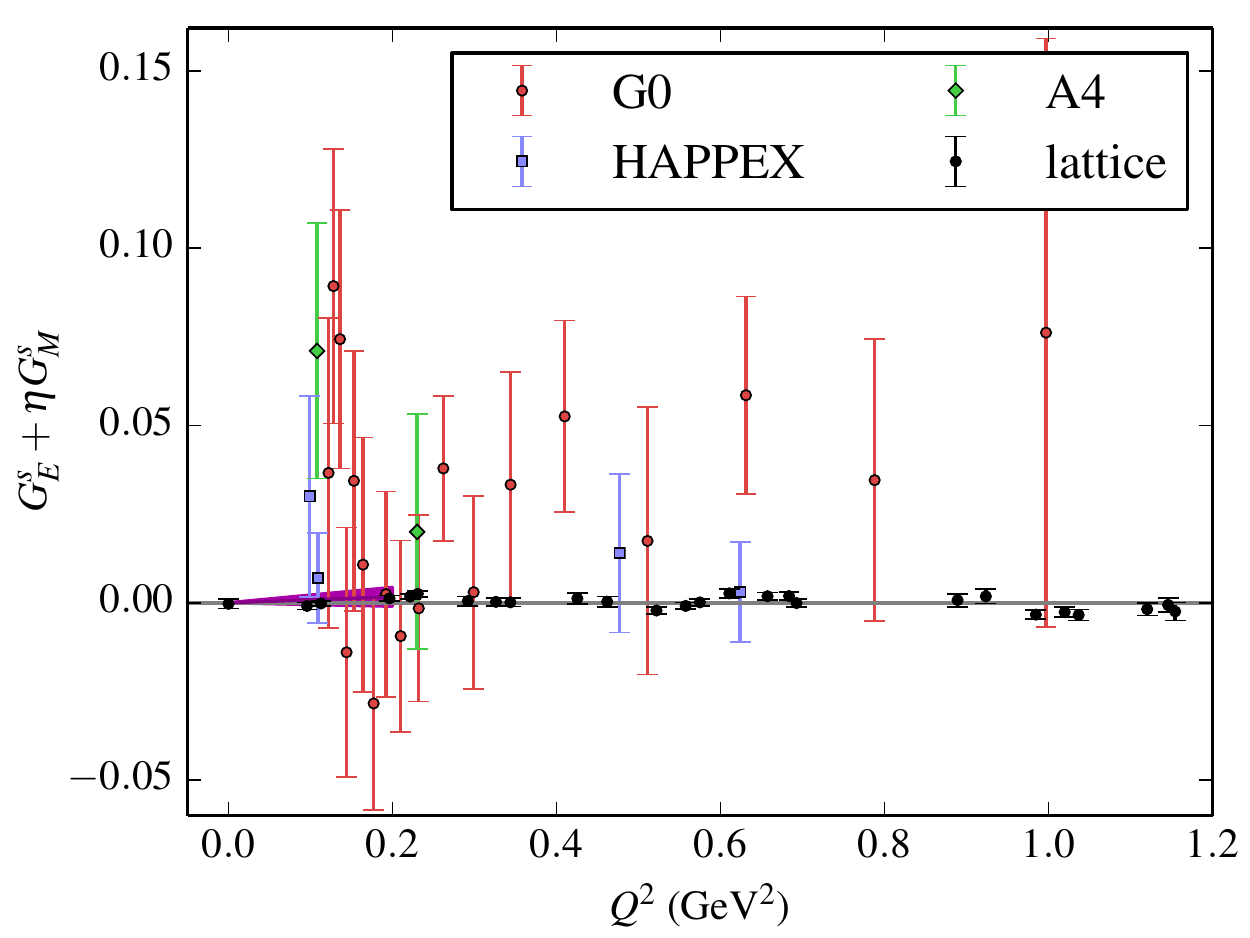}
  \caption{Linear combination of form factors, $G_E^s+\eta G_M^s$,
    probed by forward-angle parity-violating elastic $ep$ scattering
    experiments~\cite{Armstrong:2005hs, Aniol:2004hp, Aniol:2005zg,
      Acha:2006my, Ahmed:2011vp, Maas:2004ta, Maas:2004dh,
      Baunack:2009gy}.  The coefficient $\eta$ depends on the
    scattering angle and $Q^2$; for the lattice data we use the
    approximation $\eta=AQ^2$, $A=0.94\text{
      GeV}^{-2}$~\cite{Armstrong:2005hs}. In the low $Q^2$ region we
    also show the linear dependence on $Q^2$ resulting from the
    estimated charge radius and magnetic moment at the physical
    point.}
  \label{fig:expt_fwd}
\end{figure}

\begin{figure}
  \includegraphics[width=\linewidth]{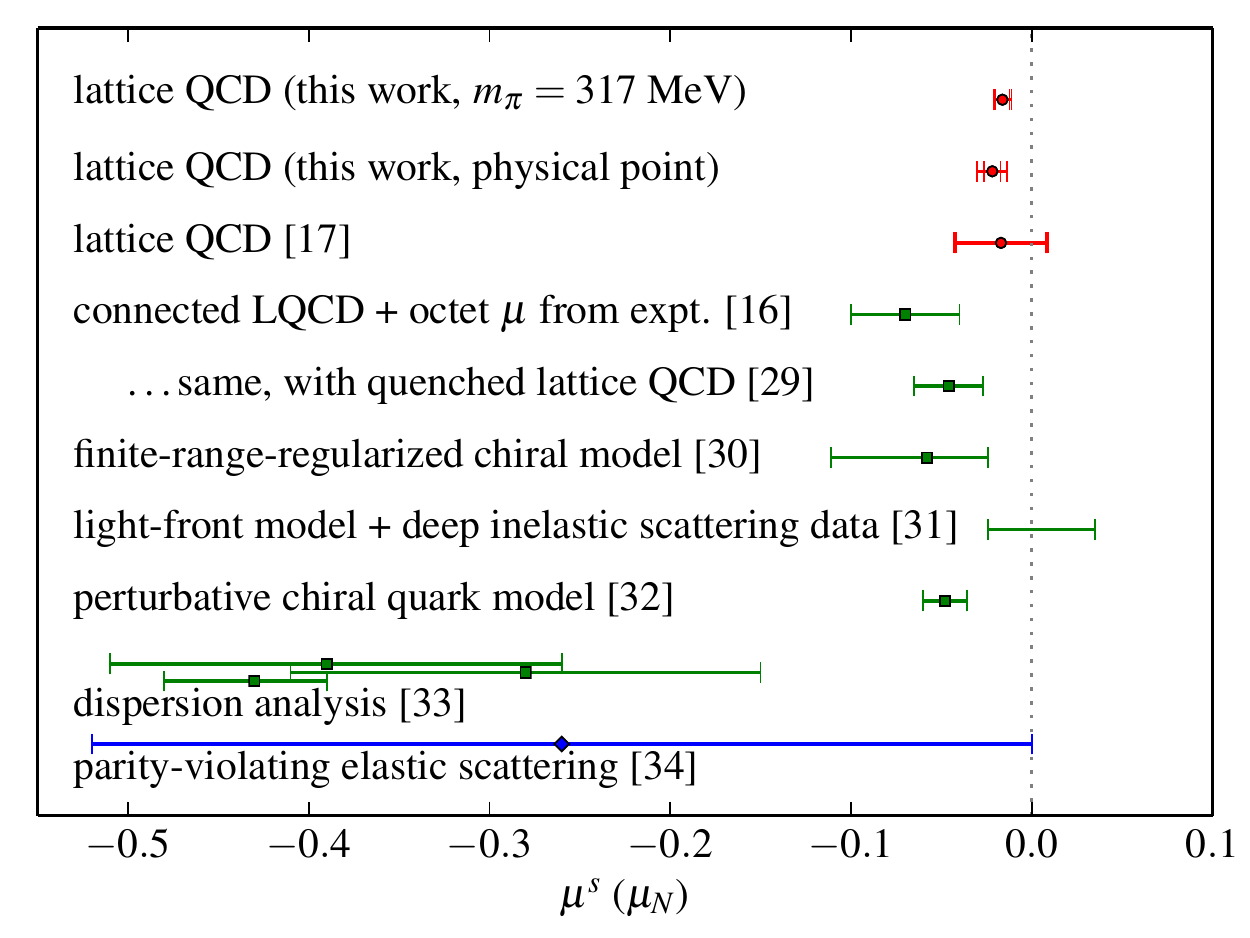}
  \caption{Determinations of the strange magnetic moment: from direct
    lattice QCD calculations (this work and Ref.~\cite{Doi:2009sq};
    red circles), models and phenomenology~\cite{Shanahan:2014tja,
      Leinweber:2004tc, Wang:2014nhf, Hobbs:2014lea,
      Lyubovitskij:2002ng, Hammer:1999uf} (green squares), and from a
    recent global analysis of parity-violating elastic scattering
    data~\cite{Gonzalez-Jimenez:2014bia} (blue diamond).}
  \label{fig:global_mus}
\end{figure}

The uncertainties are estimated as follows:
\begin{enumerate}
\item Statistical uncertainties are computed using jackknife
  resampling, using samples binned from four gauge
  configurations. These are shown as the inner error bands in
  Fig.~\ref{fig:GE_GM}.
\item Fitting uncertainties are estimated by doubling and halving the
  widths of the fit priors, as well as performing the fits using
  different estimations of correlations in the data.
\item For the light-quark disconnected form factors, we estimate
  excited-state errors by computing the form factors using a larger
  source-sink separation, 1.37~fm. We assign the same relative error
  due to excited states to the strange-quark form factors.
\item Uncertainties due to discretization effects are estimated by
  comparing against the form factors computed using the unimproved
  current.
\end{enumerate}
The outer error bands in Fig.~\ref{fig:GE_GM} show the sum of these
uncertainties in quadrature. We neglect finite-volume effects since
they are highly suppressed by $e^{-m_\pi L}$, with $m_\pi L=5.9$. Our
resulting strange radii $(r_{E,M}^2)^s\equiv -6\, \mathrm{d}G_{E,M}^{s}/\mathrm{d} Q^2|_{Q^2=0}$
and strange magnetic moment $\mu^s\equiv G_M^s(0)\,\mu_N^\text{lat}$ at pion mass 317~MeV are
the following:
\begin{equation}
  \label{eq:rE_rM_mu_317}
  \begin{aligned}
    (r_E^2)^s &= -0.0054(9)(6)(11)(2)\text{ fm}^2,\\
    (r_M^2)^s &= -0.0147(61)(28)(34)(5)\text{ fm}^2,\\
    \mu^s  &= -0.0184(45)(12)(32)(1)\:\mu_N^\text{lat},
  \end{aligned}
\end{equation}
where the four uncertainties are given in the same order as listed
above, and $\mu_N^\text{lat}$ is the nuclear magneton using the
lattice nucleon mass, 1067(8)~MeV.

Although a controlled extrapolation to the physical point would
require several lattice ensembles with varying quark masses, an
estimate can be made by combining strange-quark data with the
equivalent obtained from the disconnected light-quark form factors. By
itself, the latter can be understood in the framework of partially
quenched QCD~\cite{Bernard:1993sv} by introducing a third light quark
flavor (which couples to the current in the quark-disconnected loop)
and a bosonic ghost quark (which cancels all other loops). The
dependence of the strange radii and magnetic moment on quark masses
has been studied in SU(3) heavy-baryon chiral perturbation theory
(ChPT)~\cite{Musolf:1996zv, Hemmert:1998pi, Hemmert:1999mr}, and its
partially-quenched generalization~\cite{Chen:2001yi, Arndt:2003ww,
  Leinweber:2002qb}. At leading one-loop order, these observables
depend only on the mass $m_\text{loop}$ of a pseudoscalar meson
composed of a nucleon valence quark and a quark from the vector
current. For strange-quark and disconnected light-quark observables,
$m_\text{loop}$ is $m_K$ and $m_\pi$, respectively, and we can
interpolate to the physical kaon mass. However, using typical values
of the meson decay constant and meson-baryon couplings from
phenomenology predicts a much stronger dependence on $m_\text{loop}$
than we observe, suggesting that the quark masses are too large for
ChPT at this order. Therefore, we resort to a simple linear
interpolation in $m_\text{loop}^2$. We also adjust to the physical
nuclear magneton, and obtain at the physical point:
\begin{equation}
  \label{eq:rE_rM_mu_phys}
  \begin{aligned}
    (r_E^2)^s &= -0.0067(10)(17)(15)\text{ fm}^2,\\
    (r_M^2)^s &= -0.018(6)(5)(5)\text{ fm}^2,\\
    \mu^s &= -0.022(4)(4)(6)\:\mu_N,
  \end{aligned}
\end{equation}
where the first two uncertainties are statistical and systematic (as
estimated above). The third error is the difference between the value
at the physical point and on our lattice ensemble (using the physical
nuclear magneton), and serves as an estimate of the uncertainty due to
extrapolation to the physical point.

The experiments run at forward scattering angles were sensitive to a
particular linear combination of form factors, $G_E^s+\eta G_M^s$,
which we show in Fig.~\ref{fig:expt_fwd}. Our results and the
experimental data are both broadly consistent with zero, although the
lattice data have much smaller uncertainties. This suggests that it
will be quite challenging for future experiments to obtain a clear
nonzero strange quark signal at forward angles.

Figure~\ref{fig:global_mus} shows a comparison with some other
determinations of the strange magnetic moment. The value from
experiment has the largest uncertainty and is consistent with the
other shown results. This work is in agreement with the other values
within $2\sigma$, except for two of the dispersion-theory
scenarios~\cite{Hammer:1999uf}, and has the smallest uncertainty.

The techniques used in this work have proven effective in dealing with
the longstanding problem of noise in disconnected contributions to
matrix elements. Although future calculations at near-physical quark
masses will be needed to confirm our physical-point estimates, we have
found very small contributions from strange quarks to proton
electromagnetic observables: including the charge factor of $-1/3$
yields a roughly 0.3\% effect in the proton $r_E^2$, $\mu r_M^2$, and
$\mu$.

\begin{acknowledgments}
  \textit{Acknowledgments:} Computations for this work were carried out on
  facilities of the USQCD Collaboration, which are funded by the
  Office of Science of the U.S. Department of Energy, and on facilities provided
  by XSEDE, funded by National Science Foundation grant \#ACI-1053575.
  During this research JG, SM, JN, and AP were supported in part by
  the U.S. Department of Energy Office of Nuclear Physics under grant
  \#DE--FG02--94ER40818, ME was supported in part by DOE grant
  \#DE--FG02--96ER40965, JL was supported in part by DOE grant
  \#DE--FC02--12ER41890 and NSF grant \#CCF-121834, KO was supported in part by DOE
  grant \#DE-FG02-04ER41302 and also DOE grant \#DE-AC05-06OR23177, under which JSA operates the
  Thomas Jefferson National Accelerator Facility, SS was supported in part by DOE contract
  \#DE--AC02--05CH11231 and the RIKEN Foreign Postdoctoral Researcher
  Program, and SK was supported in part by Deutsche Forschungsgemeinschaft
  through grant SFB–TRR 55. JG was also supported in part by the PRISMA
  Cluster of Excellence at the University of Mainz, and SM was
  also supported in part by the RHIC Physics Fellow Program of the RIKEN
  BNL Research Center.
  Calculations were performed with the Chroma software
  suite~\cite{Edwards:2004sx}, using QUDA~\cite{Clark:2009wm} with
  multi-GPU support~\cite{Babich:2011:SLQ:2063384.2063478}.
\end{acknowledgments}

\providecommand{\href}[2]{#2}
\begingroup\raggedright

\endgroup

\end{document}